\documentstyle[prl,epsfig,aps]{revtex}

\def \bi{\bibitem}

 \def\(({\left(}
 \def\)){\right)}
 \def\[[{\left(}
 \def\]]{\right)}
\def\bi{\bibitem}

\newcommand {\bb}{\beta}

\def \ln{{\rm ln}}

\def \t{\tau}

\def \ln{{\rm ln}}

\def \ab2{\alpha\beta^2}

 \newcommand {\be} {\begin{equation}}
\newcommand {\bq} {\begin{eqnarray} \nonumber }
\newcommand {\ee} {\end{equation}}
\newcommand {\eq} {\end{eqnarray}}
\newcommand{\n}{\noindent}
\newcommand {\s} {\sigma}

\begin{document}
\twocolumn[\hsize\textwidth\columnwidth\hsize\csname@twocolumnfalse\endcsname
\draft       

\title{Crossover behaviour of a one dimensional Random Energy Model.}
\author{ Matteo Campellone$^{(*)}$, Silvio Franz$^{(**)}$ and
 Giorgio Parisi$^{(***)}$}
\address{
(*) Dipartimento di Fisica and Sezione INFN, Universit\`a di Padova, 
Via Marzolo 8,
 I-35131 Padova, Italy \\
 (**) International Center for Theoretical Physics\\
Strada Costiera 11,
P.O. Box 563,
34100 Trieste (Italy)\\
(***) Universit\`a di Roma ``La Sapienza''\\
Piazzale A. Moro 2, 00185 Rome (Italy)\\
e-mail:  campellone@roma1.infn.it 
franz@ictp.trieste.it, parisi@roma1.infn.it }
\date{\today}
\maketitle

\begin{abstract}
In this note we formulate a finite dimensional generalization
 of the Random Energy Model (REM) where we introduce a geometry and spatial 
 correlations between energies.  
We study the model in dimension 
one by transfer matrix techniques and we look at the crossover 
from one dimensional to mean-field behaviour.
In a first version of the model the mean field limit reproduces 
the behaviour of the original REM, while a second version of the model
 exhibits a first order phase transition with a finite latent heat.  
\end{abstract} 

\vfill

\vfill
\twocolumn
\vskip.5pc] 
\narrowtext

The Random Energy Model (REM) \cite{rem3,rem1}
 is a very simple mean field spin glass model, 
that can be exactly solved. The model, in spite of its extreme simplicity, 
captures many features of other, more complicated, spin glass models.
The REM belongs to the class of spin glasses which present a phase transition
basically due to an entropy collapse phenomenon. In other words, at low
enough temperature, the system finds no states with energies below a given 
value and remains stuck in the lowest state.
Below a critical value of the temperature the model therefore 
freezes into a state of minimal energy and zero 
entropy. This transition is rather peculiar: even though it involves no 
latent heat, it does not show any precursor effect
 (no divergent susceptibility), 
sharing therefore some features typical of first order 
transitions and others typical of second order ones.
This behaviour is qualitatively common to a whole class of mean field
 models of spin glasses (e.g. the $p>2$-spin model \cite{rem2,pspin}, 
the Potts model) \cite{elsh,grkaso} 
which show a discontinuous one-step replica symmetry breaking transition 
(1RSB) \cite{mpv}.  An interesting observation is that models of this class
 seem to be a good 
paradigm to describe structural glasses where no 
disorder is explicitly present
in the Hamiltonian but there is an effective, self-induced, randomness 
\cite{kitiwo,mapari,bume}.
The aim of this work is to provide some insight on what happens to these
models when considered in finite dimension. Some work on this line of 
research has been contemporarily 
done on a short ranged $p$-spin glass model above the lower critical 
dimension \cite{capara2,frpa,cacopa}. 
In this note we shall introduce a finite dimensional generalization
of the REM which includes a spatial dependence of the variables.
We will study in detail the properties of the model in one 
dimension where  no phase transition occurs.
Nevertheless the model is defined in such a way 
that, in the limit of a parameter $M$ to infinity, 
the mean field solution is recovered. So we will observe 
the crossover towards mean field behaviour as $M$ is increased.

The model will be formulated in two versions, which have an 
interestingly different behaviour.

In the standard REM one considers a system of $2^N$ levels with 
energies which are random independent variables extracted from a Gaussian 
distribution 

\be
P(E)\sim \exp(-E^2/N).
\label{d}
\ee

The energy levels of the system can be thought
as correspondent to the configurations of $N$ Ising spins. 

Note that we have not specified any microscopic variable for this model.
This is the natural consequence of the hypothesis that in this model 
the energies are 
totally uncorrelated from the microscopical configurations which are now only
labels of the energy level and are here indicated by the index $i$.

Usually the limit $N\to\infty $ is considered, where the system 
freezes into a state of zero entropy.
This is easily seen by the following argument. 
The average number
 of configurations 
with total energy between $E$ and $(E+\delta E )$ is
 
\be
\overline{n(E)} = 2^{N} e^{-\frac{E^{2}}{N}},
\label{ne}
\ee  
where the bar indicates the average over $P(E)$.
In the large-$N$ limit, for $|E| >E_0= N\sqrt{\ln 2}$, the entropy of 
the system is $$\ln\(( \overline{n(E)}\)).$$ 
For energies such that $|E| >E_0$ the exponent 
becomes negative and for large $N$ there are no energy levels.
 In this case the 
system is frozen into its ground state with zero entropy. 
Introducing the temperature by $$ T \equiv \(( 
\frac{\partial S(E)}{\partial E} \))^{-1} $$ 

\n
and inverting this relation one obtains the free-energy

\be
F=
\left \{
\begin{array}{ll}
-N \(( T \ln{2} + \frac{1}{4 T} \)) 
& \mbox{  for  } T> T_c = \frac{1}{ 2 \sqrt{ \ln{2} }} \\
 -N \sqrt{\ln{2}} & \mbox{  for  } T < T_c
\end{array}
\right. 
\label{F}
\ee

At the critical temperature the saddle point solution changes discontinuously
and one would say that the transition is first order. Furthermore, 
there are no physical quantities that diverge at $T_c$. Nevertheless 
the free energy is differentiable at the critical temperature 
and no latent heat is involved in the transition.

In the following we shall try to learn if the features of this 
peculiar transition are a property of the adimensional case and how 
this transition appear when the model is generalized to finite dimension.

The REM can be generalized by introducing a geometry and a
spatial correlation between the energy levels.

The Dimensional Random Energy Model (DREM)
 can be formulated in general dimension and the mean field 
solution can be found using analytical arguments.
For finite $M$ we analyze the model in $d=1$ 
by transfer matrix techniques
 and study the crossover from one dimensional to mean-field behaviour.

The DREM will be formulated in two versions, which have quite
 different mean field limit.
An interesting question is whether a growing correlation length develops or 
not for increasing $M$. 

A first version of the model shows, in the large-$M$ limit,
 a transition with no latent heat similar to the case of the REM. 
A second version of the model exhibits a first order phase transition into a
 crystalline state with a discontinuity in the specific heat.

The model is defined in the following way:
we consider a $d$-dimensional square lattice of side $L$ 
with $M$ spins on each site, in the limit of $L\to\infty$. 
So $V = L^d$ is the number of sites and $M V$ is the total number of spins
of the model.

Let us consider the link ($i \rightarrow i+\hat{\mu}$) between 
site $i$ and a nearest-neighbour site $i+\hat{\mu}$ 
where $\hat{\mu}$ is a positive unit lattice vector.

To each of the $2^{2M}$ possible configurations of the spins at the edges 
of the link $i \rightarrow i+\hat{\mu}$ we associate a random energy 
extracted from the probability
distribution

\begin{equation}
P(E(\s,\t)) = \frac{1}{\sqrt{M \pi }}
\exp{\left[\frac{-E^2 (\s,\t)}
{M}\right]}.
\label{distrdrem}
\end{equation}

 The possible energy levels of a link are therefore
$2^{2M}$ independent numbers extracted from a Gaussian distribution.   
The partition function of the model is 

\be
Z = \sum_{ \{s \} }
\exp{ \[[-\beta \sum_{i}^{V}\sum_{\hat{\mu}}
 E_{i,i+\hat{\mu}}(s_{i},s_{i+\hat{\mu}}) \]] },
\ee

\noindent
$ E_{i,i+\hat{\mu}}(s_{i},s_{i+\hat{\mu}})$ being the energy of link 
$i \rightarrow i+\hat{\mu}$. 
Note that, to avoid double counting, we for each site $i$ we consider 
only the $d$ nearest neighbours taken along the positive versus of each 
direction.
A possible version of the model 
consists in taking the energies of different links as 
independent variables, in which case we have a non translational (NTI) 
invariant spin glass. This is very similar to the short range $p$-spin
model introduced in \cite{capara2,frpa,cacopa} in the large-$p$ 
limit, where the energies are uncorrelated. 
The only difference, which should not be very
 important, is that this model does not account for the interactions
  between spins which are on the same site.  
A second possible version is a translational invariant (TI) model
\footnote{If $d>1$ the model would have to be also rotationally invariant
for our following consideration to be true.
 With the ``TI'' we will therefore mean 
also rotational invariant if $d >1$.}, 
in which the correspondence between the spin configurations and
the possible energy levels is space independent. This means that for each
 sample one assigns a law $(\sigma,\tau) \rightarrow E(\sigma,\tau)$
 extracting the values of the energy from (\ref{distrdrem}).

Both versions of the model can be formulated in a 
symmetric and non-symmetric way {\it i.e.} one can impose or not impose
the following symmetry condition which reduces by half 
the number of independent energy levels

\be
 E_{i,i+\hat{\mu}}(\s,\tau) = E_{i,i+\hat{\mu}} (\tau,\s).
\label{symm}
\ee
  
In the following we will study in detail the NTI and the symmetric
 translationally invariant model (STI).
While the first model is rather natural in a spin glass context, 
the second is more attractive in connection to the modeling 
of structural glasses. In this latter model, in fact, 
if the lattice is chessboard decomposable, 
the system has a crystalline ground state. Nevertheless, frustration 
due to the presence of the disorder, makes the 
minimization of the (free) energy a hard task, and the system 
may eventually fall in a glassy state.

Both versions of the model can be easily solved in the large-$M$
limit.
This is quite trivial in the case of the non-translationally invariant
 (NTI) DREM since one basically recovers the REM.
This is clear because for large $M$ one can consider the energies of 
the links essentially as uncorrelated so the average number
 of configurations 
with total energy between $E$ and $(E+\delta E )$ is
 
\be
\overline{n(E)} = 2^{M V} e^{-\frac{E^{2}}{MV}}.
\label{ne}
\ee 

Applying Derrida's standard 
microcanonical argument on the total energy of the system one obtains
a critical temperature of $T_{c}= 1/(2\sqrt{\ln 2})$ and a 
ground state total energy $E_{0}=-M V \sqrt{\ln 2}$. 
This is evidently the mean field solution of the model since, for large 
$M$, each spin interacts with a large number of nearest-neighbours.
The high temperature free energy density of the model is therefore 
the same as that of the REM and at 
$T_c$ the model freezes into its ground state.

The translational invariant model
behaves quite differently even at mean field level
 if formulated with the further condition of symmetry 
(\ref{symm}).

The STI-DREM has,
 in fact, a state of lower energy than $-ML\sqrt{\ln 2}$.
This can be understood by considering the possible energy levels of one 
single link.
The set of the energies of each single link is a REM with $2^{2M}$ energy levels
and has therefore a ground state energy 
$E_{i,i+\hat{\mu}}(\s_{i}^{0},\tau_{i+\hat{\mu}}^{0}) =-M\sqrt{2 \ln 2 }$.
This is true also for the NTI model or for the TI model without the symmetry
condition, but the choice of the configuration $\tau_{i+\hat{\mu}}^{0}$
on site $i+\hat{\mu}$ to minimize $E_{i,i+\hat{\mu}}$, in general 
does not allow $E_{i+\hat{\mu},i+\hat{\mu}+\hat{\nu}}$ to be minimized. 
Here $i+\hat{\mu}+\hat{\nu}$ is a general 
nearest neighbour site of $i+\hat{\mu}$.
The reason for this is that if the disorder is different
 from link to link and there is no condition assuring that the 
ground state of $E_{i+\hat{\mu},i+\hat{\nu}}$ will correspond to the 
configuration $\tau_{i+\hat{\mu}}^{0}$ on site $i+\hat{\mu}$.
 In fact the TI without the symmetry condition
 can crystallize on a 
periodic state only in those samples having $ \tilde{E_{0}} =
E(\s_{i}^{0},\s_{i+\hat{\mu}}^{0})$ so the energy of each link may
 again be in the ground state. 
 For the STI-DREM the picture is quite different: if the lattice is 
 chessboard decomposable, the spins can always arrange 
themselves in a structure that alternates in space the minimizing 
configurations $\s_{i}^{0},\tau_{i+\hat{\mu}}^{0}$. In this way, 
contrarily to the NTI model, every link is in its true ground state.
 This yields a total energy 
$\tilde{E_{0}} =-MV\sqrt{2 \ln 2 }$ which is lower than $E_{0}$.
The freezing into this true ground state will happen at a 
temperature $\tilde{T_{c}}$ at which the high temperature free energy reaches
the value $\tilde{E_{0}}$.
One has 
\be
\tilde{T_{c}} = \frac{1+\sqrt{2}}{2 \sqrt{\ln {2}}}.
\ee

This transition is first order and the latent heat is
\be
C_{lat} = \sqrt{\ln{2}}.
\ee

Below the lower critical dimension the phase transition disappears when 
$M$ is finite. We do not know at present what the critical dimension is, 
but we know that it has to be larger than one. 
However, even in dimension one,
it is interesting to study the crossover from smooth to sharp behaviour
when $M$ is increased.

For finite values of $M$ we analyzed the model in one dimension
 by transfer matrix.
For each link $i$ of the model we have a $2^{M} \times 2^{M}$ transfer 
matrix $\hat{T_{i}}$. 
For the translationally invariant model one has $$ \hat{T_{i}} \equiv 
\hat{T},$$ and it is easy to show, by standard transfer matrix arguments,
 that, in the limit of an infinite chain ($L \to \infty$),
 one can calculate the 
free energy density and the correlation length by the following identities

\bq
- \bb F  &=& \lim_{L \to \infty} \frac{1}{L} \ln (t_{1}) \equiv  
\lambda_{1}, \\
\xi &=& \[[ \ln{\left| \frac{t_{1}}{t_{2}} \right| } \]]^{(-1)},
\eq

\n
where $t_{1}$ and $t_{2}$ are respectively the first and second 
largest (in modulus) eigenvalues of $\hat{T}$ and $\lambda_{1}$ is called 
{\em maximum Lyapunov exponent} .

In the case of the NTI model one has to consider 
the product of the sequence of $L$ transfer matrices 
$ {\bf P_{L}}= \prod_{i=1}^{L} \hat{T_{i}}$ and 
define the correspondent hermitian matrix 
$${\bf V_{L}} \doteq \[[ {\bf P_{L}}^{+} {\bf P_{L}} \]]. $$

For $(L \to \infty)$ the free
energy density of the model can be calculated by making use of some 
central limit theorems for products of random matrices. 

More precisely one can say (Fustemberg theorem) that the following limit

\be
- \bb F = \tilde{\lambda_{1}}
 = \lim_{L \to \infty} \frac{1}{L}\overline{\ln || {\bf P_{L}} || }
\ee
exists with probability one.

 $ \tilde{\lambda_{1}}$ is called
 {\em maximum Lyapunov characteristic exponent} and is a 
 positive non-random quantity \cite{cripavu}.
One can define a whole set of characteristic Lyapunov exponents 

\be 
 \tilde{\lambda_{i}} 
\equiv \lim_{L \to \infty} \frac{1}{2 L} \ln (\tilde{t_{i}}), 
\ee

\n
where the $\tilde{t_{i}}$ are the eigenvalues of ${\bf V_{L}}$.
Similarly to the TI case a correlation length can then be defined 

\be
\xi = \((\tilde{\lambda_{1}} - \tilde{\lambda_{2}} \))^{(-1)}.
\ee

We computed the first and second Lyapunov characteristic exponent by 
means of the method developed by Benettin et al. \cite{cripavu}.
 
\begin{figure}[htpb]
\centerline{\epsfig{figure=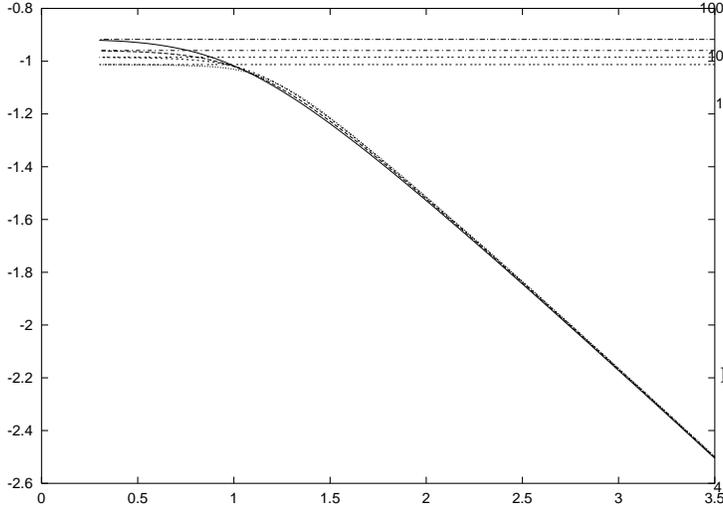,angle=270
,width=10cm}}
\vspace{0cm}
\caption[]{Free energy of the TI $d=1$ REM vs temperature for
 $M=4,5,6,7$. The horizontal lines are the ground state energies}
\label{fdrem1}
\end{figure}

\begin{figure}[htpb]
\centerline{\epsfig{figure=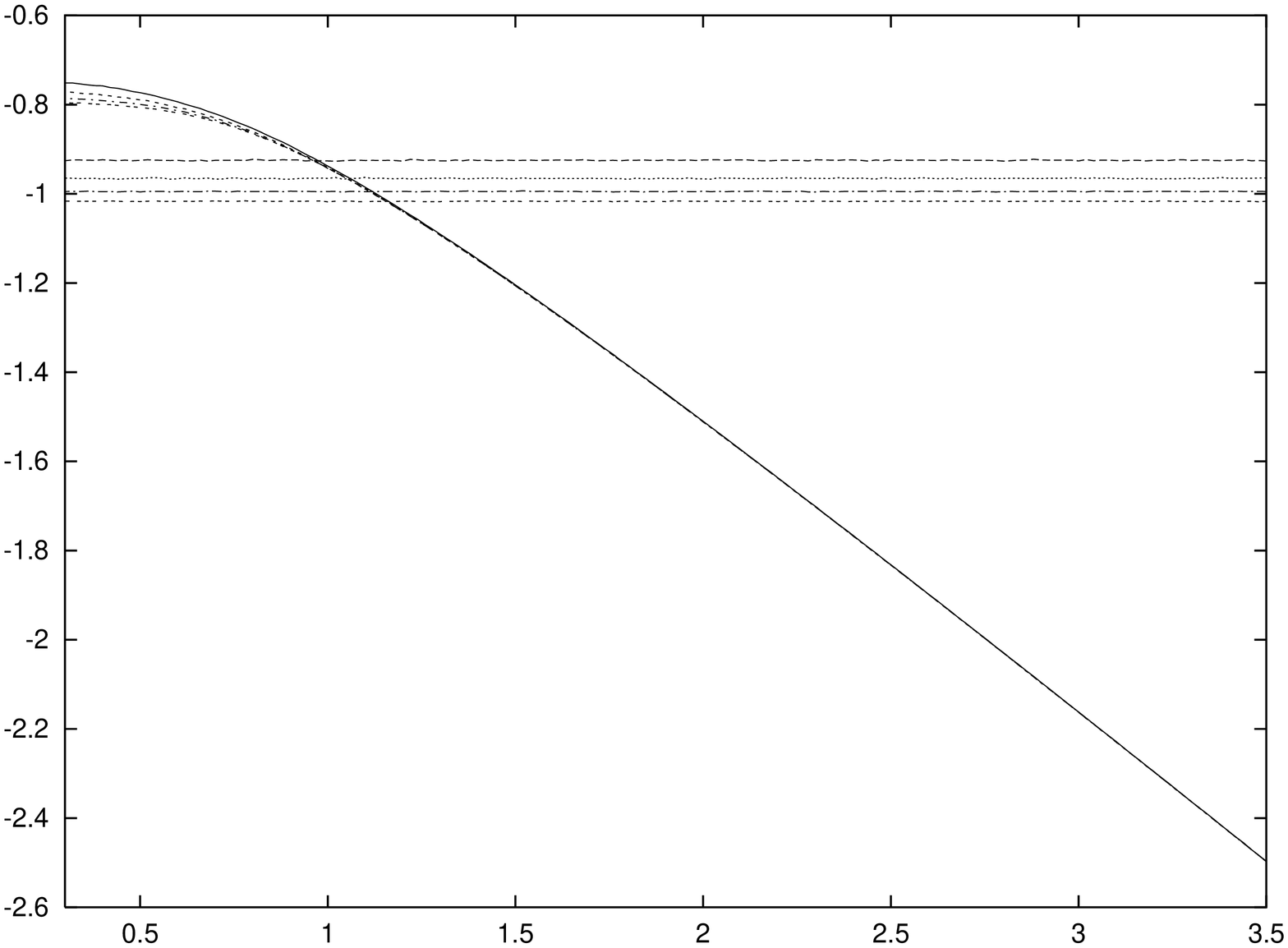,angle=270
,width=10cm}}
\vspace{0cm}
\caption[]{Free energy of the NTI $d=1$ REM vs temperature for
 $M=4,5,6,7$. The horizontal lines are the ground state energies.}
\label{fdrem2}
\end{figure}

\begin{figure}[htpb]
\centerline{\epsfig{figure=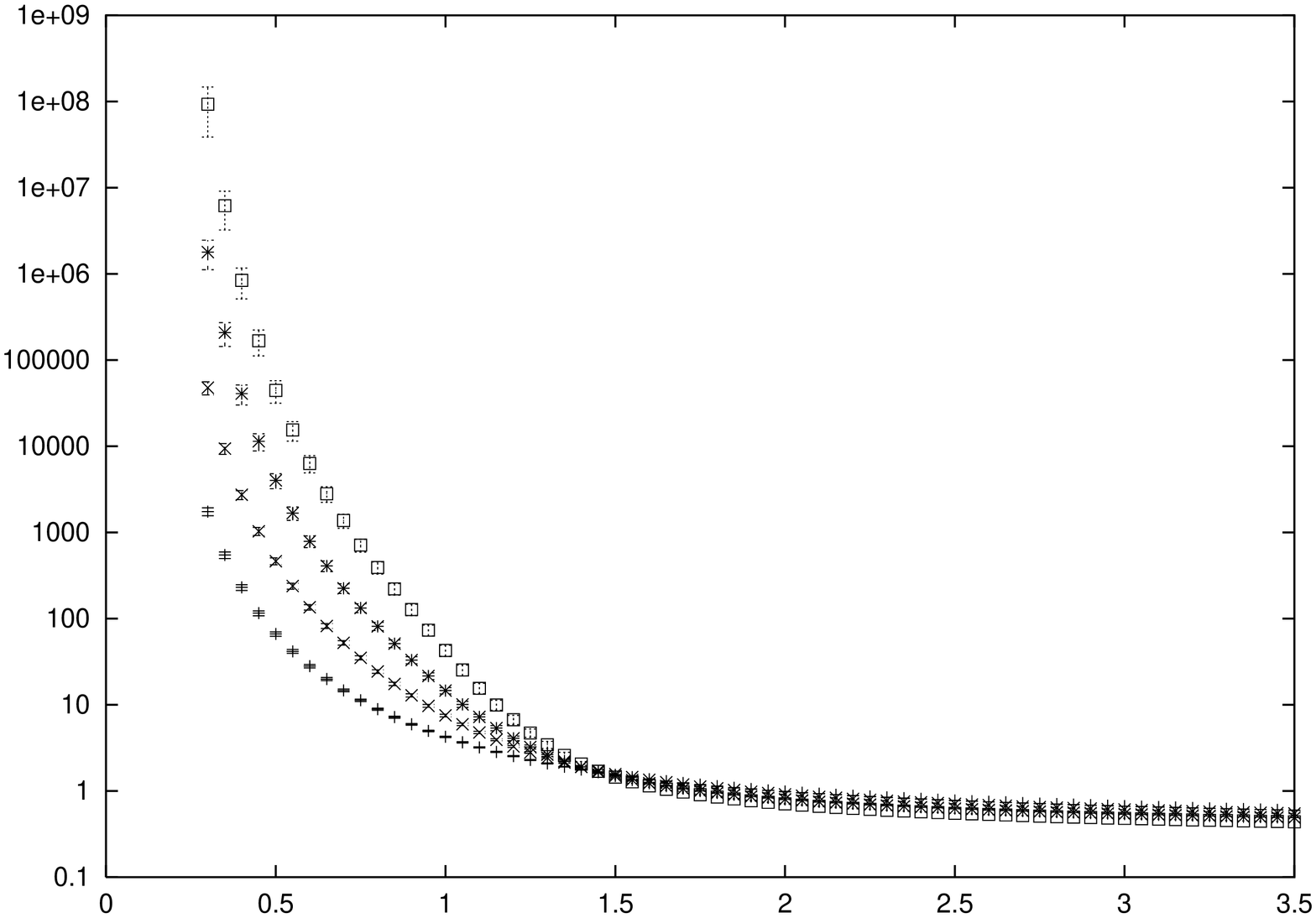,angle=270
,width=10cm}}
\vspace{0cm}
\caption[]{Correlation length of the TI $d=1$ REM vs temperature for
 $M=4,5,6,7$.}
\label{cdrem}
\end{figure}

\begin{figure}[htpb]
\centerline{\epsfig{figure=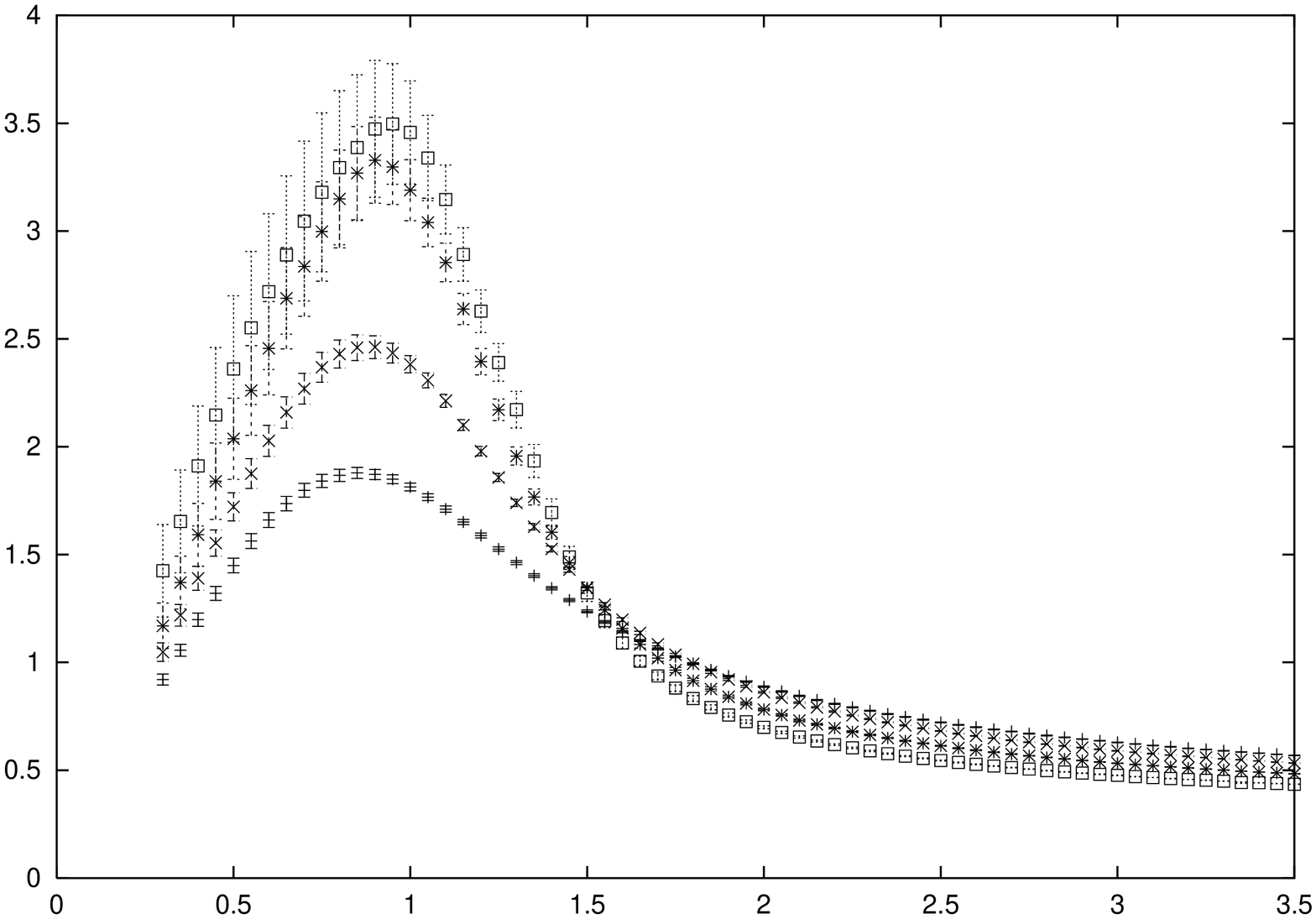,angle=270
,width=10cm}}
\vspace{0cm}
\caption[]{Ferromagnetic correlation length of the of the TI 
$d=1$ REM vs temperature for
 $M=4,5,6,7$.}
\label{cfdrem}
\end{figure}

\begin{figure}[htpb]
\centerline{\epsfig{figure=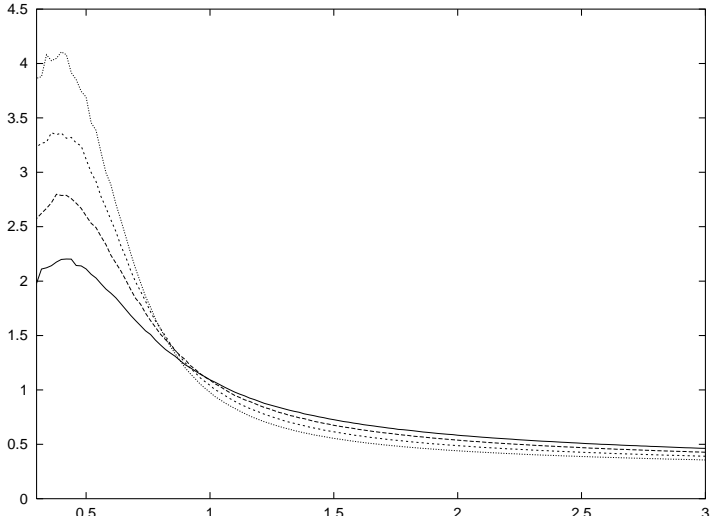,angle=270
,width=10cm}}
\vspace{0cm}
\caption[]{Correlation length of the NTI $d=1$ REM vs temperature for
 $M=4,5,6,7$.}
\label{cdremnti}
\end{figure}

 The results obtained by transfer matrix for finite $M$ are summarized in 
figure (\ref{fdrem1}-\ref{cdremnti}). The values of the free energy 
per link $F(T)$ 
are slightly different from the $M\to$ infinity 
values above but one can verify that they are consistent with them.
For finite $M$ a given sample of the TI model is more likely to freeze at 
higher temperature than the mean field value, and the free energy 
is always well above the mean field curve.

In figure (\ref{fdrem1}) and  (\ref{fdrem2}) 
 we plot the free energy density per link of the STI-DREM and of the 
NTI-DREM in function of the temperatures and for 
different values of $M$. 

In the first figure we plot the sample-averaged free energy,
 while the second needs no average in virtue of the Fustemberg theorem.
In figure (\ref{fdrem1}) the errors are of the same order of magnitude
 of the pointsize. In figure (\ref{fdrem2}) there is a small numerical 
imprecision which, 
we reckon, is responsible for the slight wiggling of all the curves for the
NTI case. 
 
We note that the non-translational invariant models as well
as the non symmetric translational invariant follow quite well
 the mean field theoretical
prediction already for quite small $M$. 
In figure (\ref{fdrem1}) and  (\ref{fdrem2}) we also potted the lowest 
energy state for different values of $M$. The TI model succeeds in freezing 
right in the lowest energy by arranging itself on the configuration of
 period 2. The NTI model does not reach its lowest energy state and freezes
 only at lower temperatures with a higher value of the energy.

In figure (\ref{cdrem}) and (\ref{cdremnti})
 we plot the correlation lengths of the models. 

For the TI model we averaged the logarithm of the correlation length 
for different samples. 
One notices that at
 $\tilde{T_{c}}$ the curves for various values of $M$ separate consistently 
and are steeper and steeper the larger $M$ is. Here one sees the effect of
 the crossover with the mean field limit since one could imagine a 
 discontinuity at $\tilde{T_{c}}$ for infinite $M$. For the TI model we 
also defined a `ferromagnetic' correlation length by computing 

\be
\xi_{f} = \[[ \ln{\left | \frac{t_{1}}{t_{f}} \right| } \]]^{(-1)},
\ee
where $t_{f}$ is the second maximum {\em positive} eigenvalue.
At low temperatures $\xi_{f}$ does not coincide with the real 
$\xi$ because there is usually a negative eigenvalue than is larger in 
modulus than $t_{f}$.

It can be seen that $\xi_{f}\equiv \xi$ only in those 
samples (that occur with frequency $O(1/M)$) 
in which the ground states happens to be on the diagonal of the matrix
 $E(\s,\tau)$. The second eigenvalue for the most
of the samples is negative because it detects an anti-ferromagnetic ordering.
The reason of this is that $ \tilde{E_{0}} = E(k_0,k_0)$ implies that 
two nearby sites tend to be have the same configuration of spins in the 
lowest energy state. The opposite happens in the anti-ferromagnetic case.
As it can be seen from figure (\ref{cfdrem}),
 the ferromagnetic correlation length does not diverge at zero temperature but
shows a peak, whose height grows with $M$, 
in correspondence of the transition temperature $\tilde{T_c}$. In figure
 (\ref{cdremnti}) we plot the correlation length of the 
 non-translationally invariant model. Consistently with our mean 
 field predictions, the peak, 
 which shows the crossover with the mean field behaviour, seems to 
 predict the MF critical temperature $T_{c}$ and not $\tilde{T_{c}}$. 
So this one-dimensional REM can give us a 
slight idea on what happens when the models that in MF present a 
discontinuous 1RSB transition are generalized to finite dimension.
 
The speculation that one could make from the results obtained in this work
 is the following: if the model is provided with 
 an underlying crystalline ground state, the transition in finite dimension 
becomes a real first order transition with a finite latent heat; 
if there is no underlying crystalline state, the transition
does not seem to show any discontinuity on the first derivative of the 
free energy as a second order transition (for the possibility of the 
arousal of divergent correlations see \cite{capara2,frpa,cacopa}).  

The numerical exact solution in dimension $d=1$ represents a 
complementary approach to what was done in \cite{capara2}
 where one started from the MF solution.
 There is still much work that has to be done on the subject. 

\vspace{.5cm}
We wish to acknowledge useful conversations with A.~Crisanti 
and A.~Vulpiani.


\begin{thebibliography}{10}


\bibitem{rem3}
B. Derrida,
\newblock Random Energy Model: Limit of a Family of Disordered Models,
\newblock {\em Physical Review Letters}, {\bf 45}(2):79-82, 1980

\bibitem{rem1}
Bernard Derrida,
\newblock Random Energy Model: an exactly solvable model of disordered systems,
\newblock {\em Physical Review B}, {\bf 24}(S):2613-2626, 1981

\bibitem{rem2}
D.J. Gross and M. Mezard,
\newblock The simplest spin glass,
\newblock {\em Nuclear Physics B 240}, {\bf FS12}:431-452, 1984

\bibitem{pspin} E. Gardner,
\newblock Spin glasses with $p$-spin interactions,
\newblock {\em  E. Marinari, G. Parisi and F. Ritort, 
J. Phys. A: Math. Gen. {\bf 27} (1994) 7615. 
J. Phys. A: Math. Gen. {\bf 27} (1994) 7647. Nuclear Physics B 257}, {\bf FS14}:747-765, 1985

\bibitem{elsh} D.~Elderfield and D.~Sherrington,
\newblock The curious case of the Potts spin glass,
\newblock {\em J.Phys.C:Solid State Phys.}, {\bf 16}:L487--L503, 1983.

\bibitem{grkaso}
D.J. Gross, I. Kanter, and H. Sompolinsky,
\newblock Mean-field theory of the Potts Glass,
\newblock {\em Physical Review}, {\bf 55}(3):304--307, 1985.

\bibitem{mpv}
For a general review on spin glasses see 
\newblock M.~Mezard, G.~Parisi, and M.A.~Virasoro,
\newblock {\em Spin Glass Theory and Beyond},
\newblock Word Scientific, 1987.


\bi{kitiwo} T.R. Kirkpatrick and D. Thirumalai, Phys.  Rev. {\bf B 36}, 
5388 (1987); T.R. Kirkpatrick and P. G. Wolynes, Phys. Rev. {\bf B 36}, 
8552 (1987);

\bi{mapari}E. Marinari, G. Parisi and F. Ritort, 
J. Phys. A: Math. Gen. {\bf 27} (1994) 7615. 
J. Phys. A: Math. Gen. {\bf 27} (1994) 7647. 

\bi{bume} J.P.~Bouchaud  M.~Mezard 
J. Phys I France, 4 (1994) 1109

\bibitem{capara2}
M.Campellone, G.Parisi and P.Ranieri
\newblock Non perturbative effects in a 
short-range $p$-spin glass model
\newblock to be published 

\bibitem{frpa}
S. Franz and G.Parisi
\newblock work in progress

\bibitem{cacopa}
M.Campellone, B.Coluzzi and G.Parisi
\newblock Numerical study of 
a short-range $p$-spin glass model in 
three dimensions.
\newblock to be published 

\bibitem{cripavu}
A. Crisanti, G. Paladin and A. Vulpiani
\newblock Products of random matrices in statistical Physics 
\newblock Springler-Verlag (1993)




\end{thebibliography}
\end{document}